# Local Collaborative Filtering: A Collaborative Filtering Method that Utilizes Local Similarities among Users

Zhaoxin Shen
School of Information Management
Wuhan University
Wuhan Hubei China
1170739963@qq.com

Dan Wu*
School of Information Management
Wuhan University
Wuhan Hubei China
danwoo@126.com

## ABSTRACT

To leverage user behavior data from the Internet more effectively in recommender systems, this paper proposes a novel collaborative filtering (CF) method called Local Collaborative Filtering (LCF). LCF utilizes local similarities among users and integrates their data using the law of large numbers (LLN), thereby improving the utilization of user behavior data. Experiments are conducted on the Steam game dataset, and the results of LCF align with real-world needs.

## CCS CONCEPTS

• Information systems → Recommender systems; Collaborative filtering.

## KEYWORDS

User Behavior Data, Recommender Systems, Collaborative Filtering, Local Similarity, Law of Large Numbers

## 1 Introduction

The vast amount of information on today's Internet makes it difficult for users to filter through it. Recommender systems aim to address this problem of information overload on the Internet. One solution is to leverage other users to help the active user with filtering. For instance, we can calculate the click-through rate (CTR) for each item across all users to reflect its popularity. However, since the method processes the entire user base, it fails to meet users' personalized needs.

User-based collaborative filtering (CF) [1] addresses this limitation. The method calculates the similarity between users based on their rating histories, and assists with filtering by leveraging users similar to the active user. However, this method suffers from the sparsity problem because it relies on there being sufficiently similar users to the active user.

To improve this, this paper proposes a CF method that utilizes *local similarities* [2][3] among users—Local Collaborative Filtering (LCF). The method assists with filtering by leveraging users whose preferences overlap with the active user's, without requiring highly similar users. Furthermore, the method is based on implicit feedback and can generate both item-item and user-item recommendations simultaneously.

This paper is organized as follows: Chapter 2 introduces the main body of our LCF method. Chapter 3 presents the experimental results. Chapter 4 conducts a discussion.

## 2 Local Collaborative Filtering (LCF)

### 2.1 Item-Item Recommendation

*2.1.1 Item-to-Item Correlation Coefficient.* For an item $i$, we wish to recommend other relevant items to users who are currently browsing it. One idea is to identify these items via users who like item $i$.

This paper employs implicit feedback, whereby user action (such as browsing, clicking or purchasing) is considered positive feedback, and inaction is considered negative feedback. Let $L(i)$ denote the set of users who provide positive feedback on item $i$. Clearly, users in set $L(i)$ generally exhibit a higher degree of preference for item $i$ than the global average level, and they also tend to prefer other related items. For the remainder of this paper, we will simply refer to the degree of preference as *preference*.

For another item $j$, according to the law of large numbers (LLN), when the number of users in set $L(i)$ is sufficiently large, their average preference for item $j$ will converge to its expectation. In this case, the average preference of $L(i)$ for $j$ can be broadly categorized into the following three scenarios:

1) Item $j$ is positively correlated with item $i$. The average preference of $L(i)$ for $j$ is higher than the global average, and the average preference increases as the positive correlation strengthens.
2) $j$ is negatively correlated with $i$. In contrast to the previous scenario, the average preference of $L(i)$ for $j$ is lower than the global average, and the average preference decreases as the negative correlation strengthens.
3) $j$ is uncorrelated with $i$. In this scenario, the preference of users in $L(i)$ for $j$ follows the population distribution, meaning their average preference for $j$ approaches the global average.

* Corresponding author.

Hence, when the number of users in set $L(i)$ is sufficiently large, their average preferences retain only those related to item $i$, eliminating irrelevant preferences. Furthermore, the average additional preference of $L(i)$ for $j$ (additional preference = preference - global average) is proportional to the correlation of $j$ to $i$.

We use the CTR of a user set for an item to reflect the average preference of the user set for the item. Thus, the average preference for $j$ across all users is represented by the CTR for $j$ across all users:

$$CTR_U(j) = \frac{|L(j)|}{|E(j)|} \tag{1}$$

Where $U$ denotes the set of all users and $E(j)$ denotes the set of users exposed [4] to item $j$. The formula represents the CTR of item $j$ among users exposed to it. Similarly, the average preference of $L(i)$ for $j$ is represented by the CTR of $L(i)$ for $j$:

$$CTR_{L(i)}(j) = \frac{|L(j) \cap L(i)|}{|E(j) \cap L(i)|} \tag{2}$$

The formula represents the CTR of item $j$ among users exposed to it in set $L(i)$. Combining this with our previous conclusion, we can define the correlation coefficient $r_i(j)$ of $j$ to $i$ as follows:

$$r_i(j) = CTR_{L(i)}(j) - CTR_U(j) \tag{3}$$

To ensure the stability of $r_i(j)$, it requires a sufficient sample size according to the LLN. In practical scenarios, the sample sizes of $CTR_{L(i)}(j)$ and $CTR_U(j)$ typically exhibit $|E(j) \cap L(i)| \ll |E(j)|$. Therefore, the stability of $r_i(j)$ primarily depends on $CTR_{L(i)}(j)$. We set $r_i(j)$ to satisfy $|E(j) \cap L(i)| > \theta_1$, where $\theta_1$ is the threshold.

Thus, by calculating and sorting the correlation coefficients of other items to item $i$, we can generate item-item recommendations for item $i$.

*2.1.2 Asymmetry of the Correlation Coefficient.* We wish to understand the relationship between $r_i(j)$ and $r_j(i)$. Since the correlation coefficient has an expectation, we may consider an extreme scenario to eliminate the influence of the exposure variable, in which the items are exposed to all users. In this case, $E(i) = E(j) = U$, so that $r_i(j)$ and $r_j(i)$ can be expressed as follows:

$$r_i(j) = \frac{|L(j) \cap L(i)|}{|L(i)|} - \frac{|L(j)|}{|U|} = \frac{|L(i) \cap L(j)||U| - |L(i)||L(j)|}{|L(i)||U|}$$
$$r_j(i) = \frac{|L(i) \cap L(j)|}{|L(j)|} - \frac{|L(i)|}{|U|} = \frac{|L(i) \cap L(j)||U| - |L(i)||L(j)|}{|L(j)||U|}$$

Since the numerators of the two equations above are identical, we can derive the ratio of $r_i(j)$ to $r_j(i)$:

$$\frac{r_i(j)}{r_j(i)} = \frac{\frac{|L(j)|}{|U|}}{\frac{|L(i)|}{|U|}} = \frac{CTR_U(j)}{CTR_U(i)} \tag{4}$$

Since $CTR_U(j)$ and $CTR_U(i)$ also have expectations, when the items receive sufficient exposure, the ratio of $r_i(j)$ to $r_j(i)$ will converge to the ratio of $E[CTR_U(j)]$ to $E[CTR_U(i)]$.

## 2.2 User-Item Recommendation

*2.2.1 User-Item Predicted Click-Through Probability.* To generate recommendations for a user $u$, we wish to predict his/her preference for each item. Since we calculated the item-to-item correlation coefficient in the previous section, we can predict the active user's preference based on his/her browsing history.

Since negative feedback in implicit feedback represents user inaction and cannot explicitly reflect user preference, we will only consider the positive feedback records of the active user. Let the items in the positive feedback records of user $u$ be denoted as $i_k$, where $k = 1, 2, \ldots, n$. For another item $j$, the preference of user $u$ for item $j$ can be broadly categorized into the following three scenarios:

1) The average correlation coefficient of item $j$ to all items $i_k$ is greater than 0. $j$ is positively correlated with many items browsed by $u$. Therefore, the preference of $u$ for $j$ is likely above the global average and increases as the average correlation coefficient rises.
2) The average correlation coefficient of $j$ to all $i_k$ is less than 0. $j$ is negatively correlated with many items browsed by $u$. Therefore, the preference of $u$ for $j$ is likely below the global average and decreases as the average correlation coefficient decreases.
3) The average correlation coefficient of $j$ to all $i_k$ approaches 0. $j$ is uncorrelated with most items browsed by $u$, so the preference of $u$ for $j$ is likely to approach the global average.

Hence, the average correlation coefficient of $j$ to all $i_k$ is approximately proportional to the additional preference of $u$ for $j$.

Since we represent the average preference of a user set for an item as the CTR of the user set for the item, we express the preference of a user for an item as the click-through probability (CTP) of the user for the item. Combining this with our previous conclusion yields the following property of the CTP of $u$ for $j$:

$$CTP_u(j) - CTR_U(j) \approx \frac{p}{n} \sum_{k=1}^{n} r_{i_k}(j) \tag{5}$$

Where $p$ denotes the *personalization coefficient*, whose value should be determined based on the application scenario. Therefore, $CTP_u(j)$ can be predicted as:

$$\widehat{CTP}_u(j) = CTR_U(j) + \frac{p}{n} \sum_{k=1}^{n} r_{i_k}(j) \tag{6}$$

It is easy to see that as $p$ increases, the level of personalization in recommendations also increases. The opposite is true when $p$ decreases.

To ensure the stability of $\widehat{CTP}_u(j)$, similarly to Section 2.1, we set it to satisfy $|E(j) \cap L(i_k)| > \theta_2$ for $k = 1, 2, \ldots, n$. Where $\theta_2$ is the threshold. Moreover, according to Chebyshev's LLN, when $n$ is sufficiently large, the average of all $r_{i_k}(j)$ converges to the average of their respective expectations $E[r_{i_k}(j)]$, where $k = 1, 2, \ldots, n$. Therefore, when $n$ is large, $\theta_2$ can be smaller than $\theta_1$,

meaning the criterion for user-item recommendation can be less stringent than that for item-item recommendation.

It is worth noting that, although Section 2.1 shows that $r_{i_k}(j)$ can be transformed into $r_j(i_k)$, this transformation restricts the sample to $L(j)$, thereby reducing the sample size. Therefore, we still use the formulation of Eq. (6) for better stability.

*2.2.2 Item-User Recommendation Probability.* Suppose we randomly select item $j$ and wish to recommend it to user $u$. It is known that the CTP of $u$ for all items follows a power-law distribution, and its probability density function (PDF) can be estimated based on $u$'s predicted CTP.

Therefore, for item $j$, we first calculate $\overline{CTP_u}(j)$, and then estimate the probability density of $\overline{CTP_u}(j)$ based on $u$'s PDF. Next, we set an appropriate *recommendation probability* to recommend $j$ to $u$, thereby regulating the distribution density of items with varying preference levels within the recommendation feed of user $u$.

## 3 Experimental Results

### 3.1 Dataset

Steam is a well-known gaming platform. Experiments[1] were conducted using a dataset[2] of user behavior on Steam. The dataset contains users' purchase records and playtime for games. We only used purchase records as implicit feedback. Due to the absence of an exposure variable, missing data was treated as negative feedback in the conventional manner. Specifically, for an item, the value is 1 for users who purchased it and 0 for those who did not. The statistical information of the dataset is shown in Table 1.

**Table 1: Statistics of the Steam dataset.**

| #user | #item | #purchase | sparsity |
|---|---|---|---|
| 12393 | 5155 | 128804 | 99.80% |

### 3.2 Item-Item Recommendation

*3.2.1 Item-Item Recommendation List.* We selected three popular games to serve as examples: The Elder Scrolls V, Dota 2, and Counter-Strike: Global Offensive (CS:GO). Their respective CTRs are 5.79%, 39.06%, and 11.39%. Table 2 shows the top 10 games with the highest correlation coefficients to these three games, whose general information is as follows:

- **The Elder Scrolls V**. Most of the list consists of first-person games with adventure elements, four of which are downloadable content (DLC) for The Elder Scrolls V. Additionally, Portal 2 and Fallout: New Vegas incorporate role-playing elements, and Sid Meier's Civilization V includes historical elements.

[1] https://github.com/zeus311333/LCF
[2] https://www.kaggle.com/datasets/tamber/steam-video-games/data

- **Dota 2**. Most of the list consists of team-based, competitive games with warfare elements, including PvP and PvE modes. Neverwinter and Magicka: Wizard Wars incorporate magical elements.
- **CS:GO**. Most of the list consists of first-person shooter (FPS) games, with four being other games in the Counter-Strike series.

Therefore, it can be seen that these items with high correlation coefficients also exhibit strong real-world relevance to the target items. Moreover, judging by their purchase volumes, most of them are popular items as well. We also observed that Dota 2, the most popular of the three target games, has the lowest correlation coefficients in the recommendation list. This may be attributed to its exceptionally large user base diluting the distinctiveness of its player demographics.

*3.2.2 Stability of the Correlation Coefficient.* As discussed in Section 2.1, the stability of $r_i(j)$ primarily depends on $CTR_{L(i)}(j)$. Therefore, we can estimate the stability of the correlation coefficient by assessing the stability of item CTR.

Since item click data is a binary variable, it can be readily verified that when the CTR is 50%, the data reaches its maximum mean absolute deviation (MAD) of 50%. To assess this scenario, we created a dataset of $1 \times 10^5$ clicks with a 50% CTR. We randomly sampled this data and calculated the mean absolute error (MAE) of the sample CTR relative to the true value across different sample sizes. Each group was sampled $3 \times 10^5$ times. The results of the experiment are shown in Figure 1.

As shown in the figure, the MAE decreases as the sample size increases, and the rate of decrease gradually slows down. Let $s$ denote the sample size. When $s \geq 16$, the MAE can be reduced to below 10%. When $s \geq 64$, the MAE can be reduced to below 5%. When $s \geq 400$, the MAE can be reduced to below 2%. When $s \geq 1,600$, the MAE can be reduced to below 1%. Based on this, we can set an appropriate value for $\theta_1$ as needed.

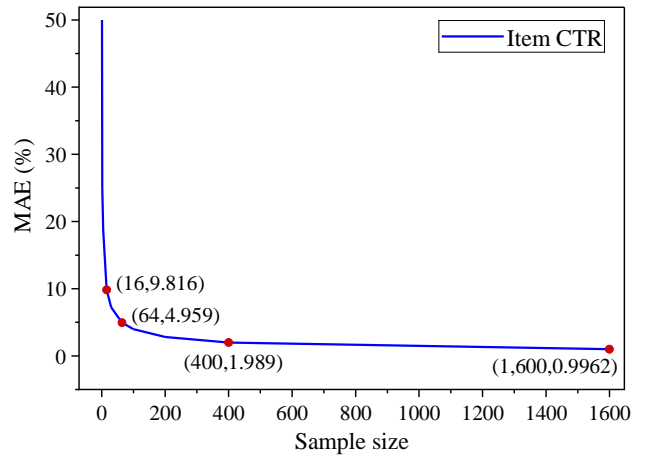

**Figure 1: Effect of sample size on item CTR stability, measured as the MAE between sample and true CTR.**

**Table 2: Item-item recommendation list for the three target games.**

| The Elder Scrolls V Skyrim | | Dota 2 | | Counter-Strike: Global Offensive | |
|---|---|---|---|---|---|
| game | $r$(%) | game | $r$(%) | game | $r$(%) |
| The Elder Scrolls V Skyrim - Dawnguard | 49.67 | Warframe | 2.77 | Team Fortress 2 | 25.94 |
| The Elder Scrolls V Skyrim - Hearthfire | 48.22 | FreeStyle2 Street Basketball | 1.98 | Left 4 Dead 2 | 24.83 |
| The Elder Scrolls V Skyrim - Dragonborn | 47.96 | Counter-Strike: Global Offensive | 1.70 | Garry's Mod | 23.85 |
| Left 4 Dead 2 | 33.19 | GunZ 2 The Second Duel | 1.67 | Unturned | 23.37 |
| Skyrim High Resolution Texture Pack | 30.88 | Dead Island Epidemic | 1.45 | Counter-Strike: Source | 23.20 |
| Borderlands 2 | 30.57 | Neverwinter | 1.37 | Counter-Strike | 17.88 |
| Portal 2 | 30.20 | Nosgoth | 1.34 | PAYDAY 2 | 17.87 |
| Counter-Strike: Global Offensive | 29.19 | Quake Live | 1.23 | Counter-Strike: Condition Zero | 16.69 |
| Sid Meier's Civilization V | 27.27 | War Thunder | 1.15 | Counter-Strike: Condition Zero Deleted Scenes | 16.69 |
| Fallout: New Vegas | 26.97 | Magicka: Wizard Wars | 1.16 | Portal 2 | 15.51 |

## 3.3 User-Item Recommendation

*Effect of the Personalization Coefficient.* The experiment employed 5-fold cross-validation, using Hit Ratio (HR) as the evaluation metric, with a recommendation list length of 10. Figure 2 illustrates the effect of the personalization coefficient $p$ on HR@10. As shown in the figure, the algorithm's performance first increases rapidly and then gradually decreases as $p$ increases, reaching its optimal value at $p = 1.5$. Notably, when $p = 0$, the algorithm exhibits zero personalization, meaning the predicted CTP of a user for an item is equivalent to the item's CTR.

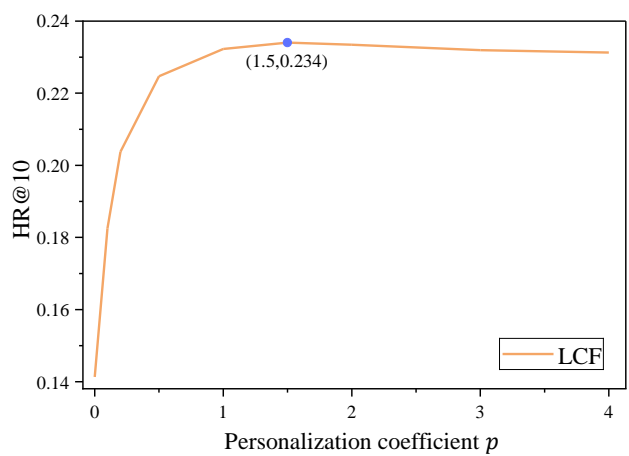

**Figure 2: Effect of personalization coefficient $p$ on LCF's performance.**

## 4 Discussion

This paper proposes a collaborative filtering method that utilizes local similarities among users. Its experimental results align with real-world needs. However, as with traditional collaborative filtering methods, this method suffers from the cold-start and sparsity problems due to the need for sufficient user behavior data to apply the LLN. Therefore, for practical applications, this method may need to be combined with other methods, such as content-based or tag-based recommendations.